\documentclass[sigconf]{acmart}

\usepackage{listings}
\usepackage{subcaption}
\lstdefinelanguage{LOGICA2}
{
  belowcaptionskip=1\baselineskip,
  breaklines=true,
  frame=topbottom,
  numbers=left,
  numberstyle=\tiny\color{gray},
  xleftmargin=0pt,
  numbersep=5pt,
  basicstyle=\small\ttfamily,
  keywordstyle=\color{black},
  commentstyle=\itshape\color{gray},
  identifierstyle=\color{black},
  backgroundcolor=\color{white},
  stringstyle=\color{black},
  showstringspaces=false,
  sensitive=false,  
  keywords={not},
  morecomment=[l]{\#},
  morestring=[b]",
  mathescape=true,
  aboveskip=4pt,
  belowskip=4pt,
}
\usepackage{xcolor}
\definecolor{steelblue}{RGB}{70, 130, 180}


\renewcommand\footnotetextcopyrightpermission[1]{}
\setcopyright{none}
\acmConference{}{}{}
\acmYear{}
\acmDOI{}
\acmPrice{}
\acmISBN{}

\title{Logical Robots: Declarative Multi-Agent Programming in Logica}

\author{Evgeny Skvortsov}
\affiliation{\institution{Google LLC}\city{Kirkland}\state{WA}\country{USA}}
\email{evgskv@google.com}

\author{Yilin Xia}
\affiliation{\institution{University of Illinois}\city{Urbana-Champaign}\state{IL}\country{USA}}
\email{yilinx2@illinois.edu}

\author{Ojaswa Garg}
\affiliation{\institution{Google LLC}\city{Kirkland}\state{WA}\country{USA}}
\email{ojaswagarg@google.com}

\author{Shawn Bowers}
\affiliation{\institution{Gonzaga University}\city{Spokane}\state{WA}\country{USA}}
\email{bowers@gonzaga.edu}

\author{Bertram Lud\"ascher}
\affiliation{\institution{University of Illinois}\city{Urbana-Champaign}\state{IL}\country{USA}}
\email{ludaesch@illinois.edu}

\begin{abstract}
We present Logical Robots, an interactive multi-agent simulation platform where autonomous robot behavior is specified declaratively in the logic programming language Logica. Robot behavior is defined by logical predicates that map observations from simulated radar arrays and shared memory to desired motor outputs. This approach allows low-level reactive control and high-level planning to coexist within a single programming environment, providing a coherent framework for exploring multi-agent robot behavior. 

\end{abstract}

\keywords{Logic Programming; Multi-Agent Systems; Declarative Programming; Robot Simulation}

\newcommand{\BibTeX}{\rm B\kern-.05em{\sc i\kern-.025em b}\kern-.08em\TeX}

\begin{document}

%%% The next command prints the information defined in the preamble.

\maketitle 

%%%%%%%%%%%%%%%%%%%%%%%%%%%%%%%%%%%%%%%%%%%%%%%%%%%%%%%%%%%%%%%%%%%%%%%%

\section{Introduction}

Logic programming has a rich history in robot planning and multi-agent systems, from approaches like Shakey and STRIPS~\cite{fikes1971strips,nilsson1984shakey} 
through situation calculus~\cite{mccarthy1969philosophical} and 
Golog~\cite{levesque1997golog}, to modern Answer Set 
Programming~\cite{erdem2018asp,son2023asp} and agent languages like Jason~\cite{bordini2007jason} and GOAL~\cite{hindriks2009goal}. While logic 
programming has been applied to reactive control (IndiGolog~\cite{de2009indigolog} and LTL synthesis~\cite{kress2009temporal}), surveys 
of logic-based MAS technologies~\cite{calegari2021logic} confirm most approaches operate at high abstraction levels, leaving low-level control to imperative code. Thus, symbolic-subsymbolic integration remains an open frontier.

We show progress on this frontier with Logical Robots\footnote{Demo website: {\color{steelblue}\url{https://logica.dev/robots}}; source code: {\color{steelblue}\url{https://github.com/EvgSkv/logica/tree/main/docs/robots}}; video: {\color{steelblue}\url{https://tinyurl.com/logicalrobots}}}, an interactive simulation platform where autonomous robots navigate labyrinths, coordinate around hazards, and pursue goals---\textbf{with symbolic planning and low-level control unified in declarative Logica programs}~\cite{skvortsov2024logica}. The key enabler is that robotics problems naturally decompose into aggregations over sensor data: from high-level reasoning such as selecting the nearest beacon (\texttt{ArgMin} over distances), to low-level control such as computing steering corrections (\texttt{WeightedAverage} over radar data). Unlike Answer Set Programming~\cite{lifschitz2019answer} and Prolog~\cite{wielemaker2012swi}, which face grounding bottlenecks with large-scale data, Logica compiles to SQL, treating sensor streams as relational tables and performing aggregations at modern database speeds. This enables sensor fusion, reactive control, and symbolic planning within a single coherent framework.

\textbf{Contributions:} This work provides: (1) A demonstration that Logica can help unify symbolic planning and low-level control through declarative aggregations that process large scale sensor data; and (2) An interactive multi-agent simulation platform with ten examples of progressively challenging coordination scenarios.

%%%%%%%%%%%%%%%%%%%%%%%%%%%%%%%%%%%%%%%%%%%%%%%%%%%%%%%%%%%%%%%%%%%%%%%%
\section{The Logical Robots Platform}
\subsection{Simulation Ontology}
\begin{figure}[t]
  \centering
  \includegraphics[height=4cm]{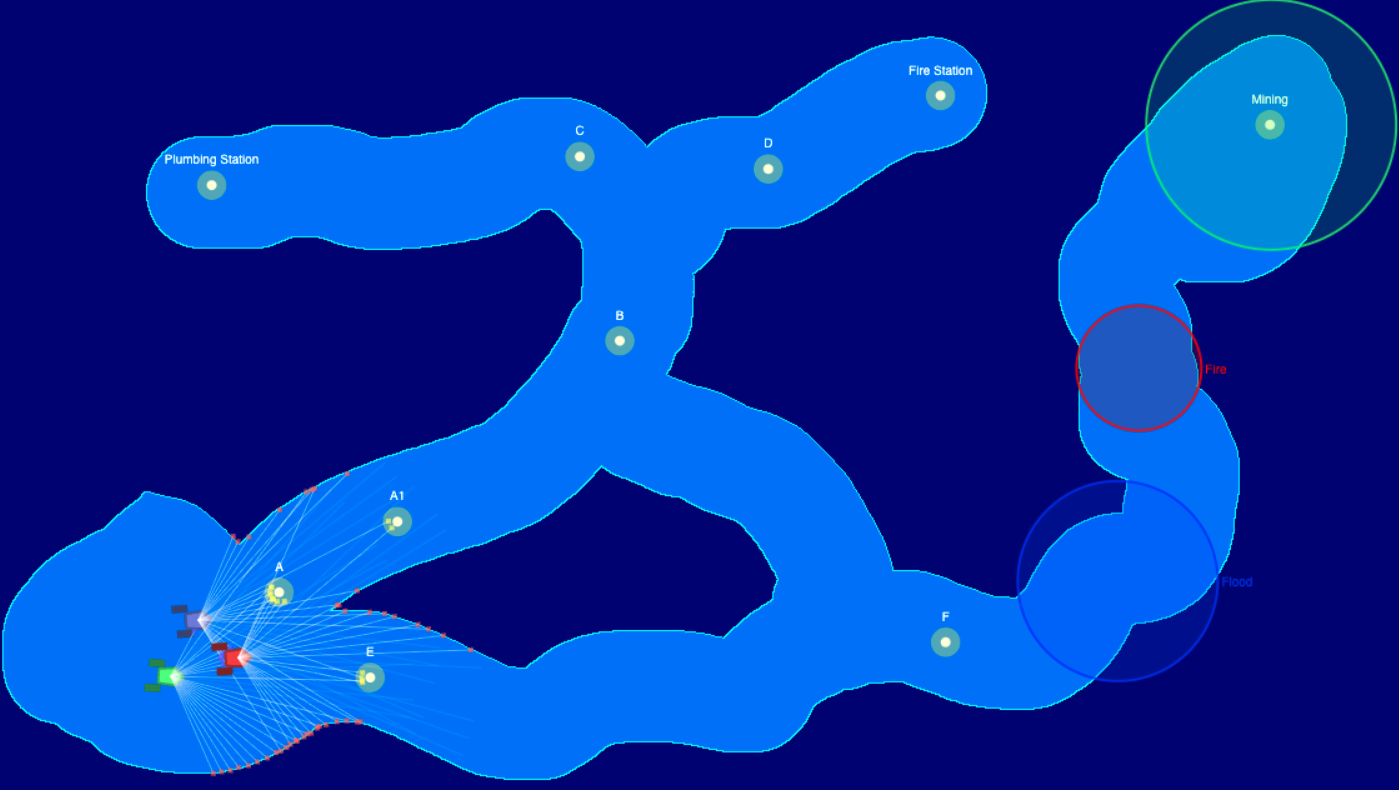}
  \caption{Multi-robot exploration with sensor rays (red endpoints: wall, green endpoints: beacon) and dynamic area toggling triggered by specific beacon detection (e.g., Fire Station).}
  \label{fig:areas}
\end{figure}

Logical Robots provides a two-dimensional labyrinth simulation environment (Figures~\ref{fig:areas} and \ref{fig:labyrinth}) containing four entity types:

\begin{itemize}
    \item \textbf{Robots} act as autonomous agents (colored squares with radiating sensor lines), each independently executing the same Logica program for movement control. Each robot is identified by a unique name.
    
    \item \textbf{Beacons} define stationary waypoints (e.g., A, B, Fire Station) serving as navigational aids and area triggers.
    
    \item \textbf{Areas} are spatial regions marked by colored circles (blue and red) that toggle between accessible and restricted states when robots detect designated beacons.
    
    \item \textbf{Win Conditions} represent success criteria,  requiring robots to reach target zones (e.g., Mining zones) simultaneously.
\end{itemize}

% \subsection{Perceptual Interface and Observations}
\vspace{-3pt}
\subsection{Sensing: Sensors and Memory}
Each default labyrinth comes with a fixed number of robots with initial positions. The platform provides two input predicates for robots to process and decide their next move:

\begin{itemize}
  \item \texttt{Sensor(robot\_name:, sensor:)} provides immediate observations: \texttt{sensor.radar} returns an array of sensor rays, each containing \texttt{angle}, \texttt{distance}, \texttt{object} type (beacon/wall/robot/none), and \texttt{label} (beacon ID or robot name) values. 

 \item \texttt{Memory(robot\_name:, memory:)} accesses stored information in user-defined data structures (e.g., strings, lists, JSON). By default, each robot accesses only its own memory. However, users can also configure the predicate to allow access to the memory of the other robots in the simulation.
 % others'. 
\end{itemize}

\vspace{-6pt}
\subsection{Reactive Control Logic}

Robot behavior is defined by a \texttt{Robot} predicate that specifies the \textbf{execution model}. The simulation operates in discrete synchronous rounds: each robot independently executes its Logica program once per timestep, reading \texttt{Sensor} and \texttt{Memory} predicates as inputs and producing \texttt{desire} and updated \texttt{memory} as outputs. Each robot maintains its own memory store that persists across timesteps until overwritten, with user-configurable cross-robot read access. A complete obstacle-avoiding robot is defined in the following program.

\begin{lstlisting}[language=LOGICA2]
FreedomMotion(radar) = WeightedAverage {
  x.distance -> x.angle :- x in radar
};
Robot(robot_name:, desire:, memory: "I am a robot") :-
  Sensor(robot_name:, sensor:),
  freedom = FreedomMotion(sensor.radar),
  speed = 0.5,
  desire = {
    left_engine:  speed - freedom + 0.1,
    right_engine: speed + freedom
  };
\end{lstlisting}

\noindent 
In this simple example, Line 1 defines \texttt{FreedomMotion} as a weighted average where each radar ray votes for its angle, weighted by its distance value. Rays detecting distant objects (or no obstacles) cast stronger votes than rays hitting nearby obstacles, 
which directs the robot to open space.
% effectively computing directions with the most open space. 
Lines 4--11 define the robot's behavior: it reads sensor data (line 5), computes the preferred direction (line 6), and converts this into differential drive commands (lines 9--10). Positive \texttt{freedom} slows the left engine and speeds up the right (turning left), while negative \texttt{freedom} does the opposite. The 0.1 asymmetry breaks deadlocks in symmetric configurations.

% For iterative algorithms like distributed pathfinding, each simulation tick performs one iteration by reading the current \texttt{Memory} and writing updated estimates.
\vspace{-3pt}
\subsection{Planning With State}
\begin{figure}[t]
  \centering
  \includegraphics[height=4cm]{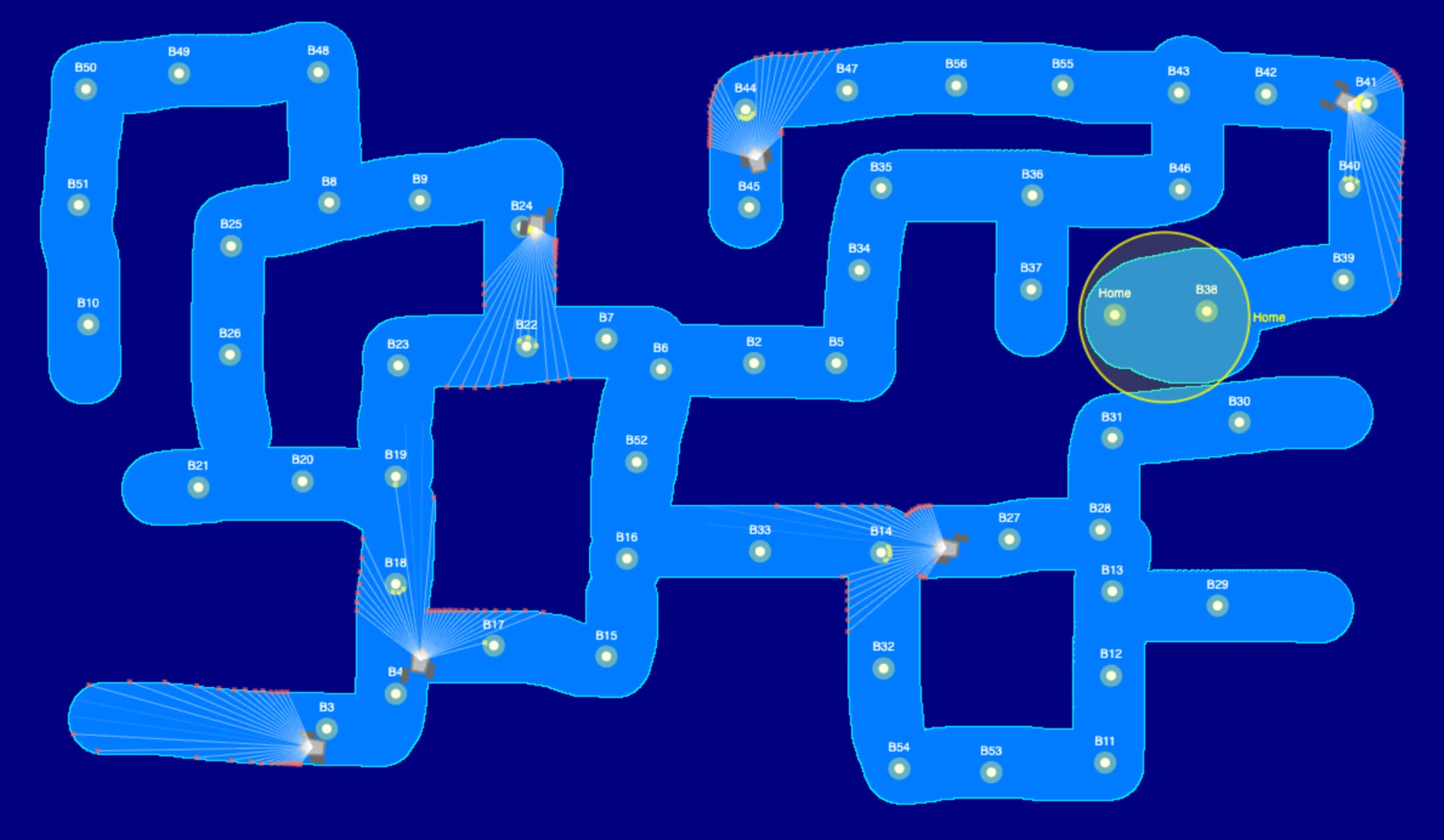}
  \caption{Distributed mapping scenario: robots collaboratively explore a beacon-filled labyrinth, building a shared navigation graph to find their way ``Home.''}
  \label{fig:labyrinth}
\end{figure}

While obstacle avoidance is purely reactive, more complex behaviors require state. Consider the distributed pathfinding solution for Figure~\ref{fig:labyrinth}: robots independently observe beacons and store pairwise distances in local memory, while a leader robot aggregates all of the memories of the robots to build the beacon network. Once any robot reaches ``Home'', the leader computes shortest paths, enabling other robots to navigate via visited beacons. 

The following Logica rule computes, as part of the Bellman-Ford pathfinding algorithm, the \verb|PosteriorHomeDistance|, which iteratively updates each beacon's shortest distance to ``Home''. The \verb|Min=| aggregation evaluates three cases and selects the minimum: (1) keep the previous distance \verb|HomeDistance(beacon)|; (2) set distance to zero if this is the ``Home'' beacon; or (3) compute distance via a neighbor as \verb|HomeDistance(neighbor) + D(neighbor,beacon)|, where \verb|D| represents edge distances between beacons (computed from the radar observations of the robots). Over multiple timesteps, the shortest-path information is recursively propagated through the beacon network.

\pagebreak

\begin{lstlisting}[language=LOGICA2]
PosteriorHomeDistance(beacon) Min= d :-
  d == HomeDistance(beacon) |
  d == 0, beacon == "Home"  |
  d == HomeDistance(neighbor) + D(neighbour, beacon);

\end{lstlisting}

%%%%%%%%%%%%%%%%%%%%%%%%%%%%%%%%%%%%%%%%%%%%%%%%%%%%%%%%%%%%%%%%%%%%%%%%
\vspace{-3pt}
\section{Demonstration Plan}
We will first introduce the platform interface and demonstrate the core functionalities of Logical Robot: loading labyrinths, writing robot control programs in Logica, executing simulations, utilizing the debug mode and customizing labyrinths with live editor.
Next, we will demonstrate three of the ten example scenarios in detail to highlight the platform's capabilities:
% Following the interface introduction, we will demonstrate three scenarios from the ten progressively challenging levels (Levels 7, 8, and 10) that highlight the platform's capabilities:

\begin{enumerate}

\item \textbf{Level 7: Station Management.} Robots have different goals, where some maintain contact with station beacons to disable hazards, while others navigate to the Mining area (Figure~\ref{fig:areas}). 
% This demonstrates how robots with distinct objectives can coordinate.
% This requires implicit coordination: robots autonomously distribute across station beacons to maintain safe passage for one robot to reach the Mining goal.
\item \textbf{Level 8: Formation Navigation.} Robots read the memory of encountered robots to follow them and eventually reach the home area together as a group.
% The challenge demonstrates how local sensor-based rules can produce global coordination.

\item \textbf{Level 10: Distributed Mapping.} Robots collaboratively discover beacon positions. A leader robot maintains shared memory aggregating these discoveries, enabling distributed Bellman-Ford pathfinding where robots follow computed paths through the beacon network.
\end{enumerate}

\noindent
The platform's increasingly complex example levels, and the ability to create new levels, positions Logica Robot as a novel educational tool for learning logic-based programming techniques and their use in declaratively building multi-agent systems.

%%%%%%%%%%%%%%%%%%%%%%%%%%%%%%%%%%%%%%%%%%%%%%%%%%%%%%%%%%%%%%%%%%%%%%%%
% \section{Conclusions}
% Logical Robots demonstrates that declarative logic programming is viable for multi-agent control, with planning and low-level reactive behaviors unified through declarative aggregations. The platform serves both as a proof-of-concept for logic-based robot coordination and as an educational tool for teaching multi-agent programming through progressively challenging scenarios.
%%%%%%%%%%%%%%%%%%%%%%%%%%%%%%%%%%%%%%%%%%%%%%%%%%%%%%%%%%%%%%%%%%%%%%%%

% \begin{acks}
% If you wish to include any acknowledgments in your paper (e.g., to 
% people or funding agencies), please do so using the `\texttt{acks}' 
% environment. Note that the text of your acknowledgments will be omitted
% if you compile your document with the `\texttt{anonymous}' option.
% \end{acks}

%%%%%%%%%%%%%%%%%%%%%%%%%%%%%%%%%%%%%%%%%%%%%%%%%%%%%%%%%%%%%%%%%%%%%%%%

\bibliographystyle{ACM-Reference-Format} 
\bibliography{aamas26}

%%%%%%%%%%%%%%%%%%%%%%%%%%%%%%%%%%%%%%%%%%%%%%%%%%%%%%%%%%%%%%%%%%%%%%%%

\end{document}